\documentstyle[aps,preprint]{revtex}
\begin{document}
\draft
\title{
Bosonization of One-Dimensional Exclusons and
Characterization of Luttinger Liquids
}
\author{Yong-Shi Wu and Yue Yu}
\address{Department of Physics, University of Utah,
Salt Lake City, UT 84112}
\maketitle
\begin{abstract}
We achieve a bosonization of one-dimensional
ideal gas of exclusion statistics $\lambda$ at low
temperatures, resulting in a new variant of $c=1$
conformal field theory with compactified radius
$R=\sqrt{1/\lambda}$. These ideal excluson gases
exactly reproduce the low-$T$ critical properties
of Luttinger liquids, so they can be used to
characterize the fixed points of the latter.
Generalized ideal gases with mutual statistics
and non-ideal gases with Luttinger-type interactions
have also similar behavior, controlled by an effective
statistics varying in a fixed-point line.

\end{abstract}



Recently a new combinatoric rule for many-body state
counting\cite{Hald1,Wu}, which essentially is an abstraction
and generalization of Yang-Yang's counting\cite{YangYang,BerWu}
in Bethe ansatz solvable models, is shown to be applicable to
elementary excitations in a number of exactly solvable models
for strongly correlated systems [1,2,4-7] in one, two
and higher dimensions. This has led to the notion of
fractional exclusion statistics\cite{Hald1} (FES) and associated
generalized ideal gases\cite{Wu} (GIG). (For abbreviation, we
will call particles or excitations obeying FES as {\it
exclusons}.)  An important issue to be addressed
is the relevance of FES to realistic gapless
systems for which state counting is somewhat obscure
due to strong correlations among particles. In this Letter,
we suggest that at least {\it for some
strongly correlated systems or non-Fermi liquids,
their low-energy or low-temperature fixed point may be
described by a GIG associated with FES}, similar to the
way that of Landau-Fermi liquids by ideal Fermi gas\cite{RG}.

A well-established class of non-Fermi liquids is the Luttinger
liquids in 1D, proposed by Haldane\cite{Hald2}. To produce
in this case a testimony to our above suggestion, we will show
that the low-$T$ critical properties of Luttinger liquids are
exactly reproduced by those of 1D ideal excluson gases (IEG),
if one identifies the controlling parameter\cite{Hald2}
of the former with the statistics $\lambda$ of the latter.
So {\it IEG can be used to describe the fixed points of
Luttinger liquids}. A main development here is that we have
succeeded in bosonizing 1D excluson systems at low $T$,
{\it \`a la} Tomonaga\cite{Tomonaga}. It results in a new
variant of conformal field theory (CFT) with central charge
$c=1$ and compactified radius \cite{CFT} $R=\sqrt{1/\lambda}$.
The particle-hole duality between $\lambda$ and $1/\lambda$
in IEG \cite{BerWu,NaWil} gives rise to a duality between $R$
and $1/R$ in this variant of $c=1$ CFT.

We have also studied the effects of mutual
statistics between different momenta and
Luttinger-type (density-density) interactions
among exclusons. In both cases, the low-$T$
behavior is controlled by an effective statistics
$\lambda_{eff}$ for excitations near the
Fermi points, the same way that of IEG by $\lambda$.
In 1D both the momentum independent part of interactions
and change in chemical potential $\mu$ are {\it relevant}
perturbations\cite{RG,Schulz}, resulting in a continuous
shift in the fixed-point line parametrized by $\lambda$.

Consider a GIG of $N_{0}$ particles on a ring
with size $L$. Single-particle states are labeled
by $k_i$. The total energy and momentum
are given by $E=\sum k^2_i$ and $P=\sum k_i$.
We assume \cite{Hald1,Wu} that in the thermodynamic
limit the density, $\rho_a (k)$, of available
single-particle states is {\it linearly}
dependent on the particle density, $\rho(k)$.
By definition, the statistics matrix is given by
the derivative
  \begin{equation}
g(k,k')= - \, \delta\rho_a(k)/ \delta\rho(k').
\label{stat}
  \end{equation}
The system is called an IEG of statistics $\lambda$
(with {\it no mutual statistics} between
different momenta), if $g(k,k')=\lambda\, \delta(k-k')$,
or $\rho_a(k)+\lambda\rho(k)=\rho_{0}(k)$ \cite{BerWu},
where $\rho_0(k)\equiv 1/2\pi$ is the bare density of
single-particle states. Thus, $\lambda=1$ corresponds
to fermions, and $\lambda=0$ to bosons.
The thermodynamics of IEG is shown \cite{Wu}
to be determined by the thermodynamic potential
 \begin{equation}
\Omega=-\frac{T}{2\pi}\int_{-\infty}^\infty dk
\,\ln(1+w(k,T)^{-1}),
\label{Omega}
\end{equation}
with the function $w(k,T)\equiv \rho_{a}(k)/\rho(k)$
satisfying
\begin{equation}
w(k,T)^{\lambda} [1+w(k,T)]^{1-\lambda}=e^{(k^2-\mu)/T}.
\label{forW}
\end{equation}
In the ground state, there is a Fermi surface
such that $\rho(k)=1/2\pi\lambda$ for $|k|<k_{F}$
and $\rho(k)=0$ for $|k|<k_{F}$. Then the Fermi momentum is
given by $k_F=\pi\lambda \bar{d}_{0}$, and the ground
state energy and momentum by $E_0/L=\pi^2\lambda^2
\bar{d}_{0}^{3}/3$, $P_{0}=0$, with the average density
$\bar{d}_{0}=N_{0}/L$.

Now let us examine possible excitations in the IEG\@.
First there are density fluctuations due to
particle-hole excitations, i.e.\ sound waves with
velocity $v_{s} =v_{F}\equiv 2k_{F}$ (see below).
Moreover, by adding extra $M$ particles to the
ground state, one can create particle excitations,
and by Galileo boost a persistent current. Our
observation is that the velocities of these three
classes of elementary excitations in IEG satisfy
a fundamental relation
that Haldane years ago used to characterize the
Luttinger liquid\cite{Hald2}.
Indeed, shifting $N_0$ to $N=N_0+M$, the
change in the ground state energy is $\delta_1 E_0=
\pi(\lambda k_F)M^2$, while a persistent current,
created by the boost of the Fermi sea $k\to k+\pi J/L$,
leads to the energy shift
$\delta_2E_0 =\pi(k_F/\lambda)J^2$. (Due to periodic
boundary conditions, $M$ and $J$ are constrained
\cite{Hald2} by $M=J\,({\rm mod}\; 2)$.)
Therefore the total change in energy and in momentum,
due to charge and current excitations are
\begin{equation}
\delta E_0=\frac{\pi}{2}(v_N M^2+v_J J^2),~~
\delta P_0=\pi(\bar{d}_0+\frac{M}{L})J,
\end{equation}
respectively, with
 \begin{equation}
v_N=v_s\lambda, \;\; v_J=v_{s}/\lambda, \;\;
v_{s} = \sqrt{v_{N}v_{J}}.
\label{Velo}
 \end{equation}
These coincide with the well-known relations\cite{Hald2}
in Luttinger liquid theory, if we identify $\lambda$
with Haldane's controlling parameter $\exp(-2\varphi)$.
So it is interesting to see whether all critical exponents
of Luttinger liquids would be reproduced simply by IEG\@.

To calculate the exponents in IEG, we need
to develop a bosonization approach. Following Yang
and Yang\cite{YangYang,Suth}, we introduce the dressed
energy $\epsilon (k,T)$ by writing
 \begin{equation}
w(k,T)=e^{(\epsilon (k,T)-\mu)/T}.
\label{dressE}
 \end{equation}
The point is that the grand partition function $Z_G$,
corresponding to the thermodynamic potential (2),
is of the form of that for an ideal system of fermions
with a complicated, $T$-dependent energy dispersion
given by the dressed energy:
$Z_G=\prod_k(1+e^{\mu-\epsilon (k,T)/T})$.
However, this fermion representation is not very useful,
because of the implicit $T$-dependence of the dressed
energy. We observe, nevertheless, that in the low-T
limit, the $T$-dependence of $\epsilon (k,T)$ can be
ignored: according to eqs. (3) and (6),
$\epsilon (k,T)=\epsilon (k)+O(e^{-|\epsilon |/T})$, where
\begin{equation}
\epsilon(k)=\Biggl\{ {\begin{array}{ll}
  (k^2-k^2_F)/\lambda +k_F^2, & |k|<k_F, \\
     \;\; k^2,&|k|>k_F.
 \end{array}}
\end{equation}
Thus, the low-$T$ grand partition function can be
obtained  from
the effective Hamiltonian given by
$H_{eff}=\sum_k \epsilon(k) \;c_k^\dagger c_k,$
where $c_k^\dagger$ are fermionic creation operators.

Another simplification in the low-$T$ limit
is that we need to consider only low-energy
excitations near Fermi points $k\sim \pm k_F$,
where the left- and right-moving sectors are
separable and decoupled: $H_{eff}=H_{+} + H_{-}$,
and $H_{\pm}$ has a linearized energy dispersion
\begin{equation}
\epsilon_\pm (k)=\Biggl\{  {\begin{array}{ll}
\pm v_F(k\mp k_F)+k_F^2, \;\;\; &|k|>k_F,\\
\pm v_F (k\mp k_F)/\lambda+k_F^2, &|k|<k_F.\\
\end{array}}
\end{equation}
We note the `refractions' at $k=\pm k_F$.
In spite of this peculiarity, we have succeeded
in bosonizing the effective Hamiltonian as follows.
The density fluctuation operator at $k\sim k_F$ is
constructed as follows
\begin{equation}
\begin{array}{rcl}
\rho_q^{(+)}&=&\displaystyle \sum_{k>k_F}:c^\dagger_{k-q}c_k:
+ \displaystyle \sum_{k<k_F-\lambda q}
:c^\dagger_{k+\lambda q}c_k:\\
& &\;\;\;\;\; + \displaystyle
\sum_{k_F-\lambda q< k < k_F}:c^\dagger_{\frac{k-k_F}
{\lambda}+k_F+q}c_k:
\end{array}\\
\label{cdensity}
\end{equation}
for $q>0$.
A similar density operator $\rho_q^{(-)}$ can also
be defined for $k\sim -k_F$.
Within the Tomonaga approximation \cite{comm1}, in which
commutators are taken to be their ground-state expectation
value, we obtain
\begin{equation}
[\rho_q^{(\pm)},\rho_q^{(\pm)\dagger}]=
\frac{\lambda L}{2\pi} q\, \delta_{q,q'},\;\
[H_{\pm},\rho_q^{(\pm)}]= \pm v_Fq\rho_q.
\label{density}
\end{equation}
which describe 1D free phonons with the sound
velocity $v_{s}=v_F$. Introducing normalized
boson annihilation operators $b_q=
\sqrt{2\pi/\lambda qL}\,\rho_q^{(+)}$,
$\tilde{b}_q=\sqrt{2\pi/\lambda qL}\,
{\rho}_q^{(-)\dagger}$, the bosonized
Hamiltonian satisfying (\ref{density}) is given by
\begin{equation}
H_B=v_s\{ \sum_{q>0}q(b_q^\dagger b_q
+\tilde{b}_q^\dagger \tilde{b}_q)
+\frac{1}{2}\frac{\pi}{L}[\lambda M^2
+\frac{1}{\lambda} J^2] \}.
\label{bosonH}
\end{equation}

The construction of the bosonized momentum operator
is a bit more tricky. Each term in eq. (9)
should carry same momentum $q$, therefore the fermion
created by $c_{k}^\dagger$ carries a dressed momentum $p$,
which is related to $k$ by
\begin{eqnarray}
p(k)=\Biggl\{ {\begin{array}{lll}
 k-k_F+ (k_F/\lambda),\;\;\; & k>k_F, \\
 \;\; k/\lambda,  & |k|<k_F,.\\
 k+k_F- (k_F/\lambda), & k<-k_F.\\
\end{array}}
\label{dressP}
\end{eqnarray}
In terms of this variable, the linearized dressed energy
$\epsilon (p)$ is of a simple form:
$\epsilon_{\pm}(p)=\pm v_s(p\mp p_F)+\mu$, with
$p_F=k_F/\lambda$. The bosonized total momentum
operator, corresponding to the fermionized
$P=\sum_{k} p(k)\, c_{k}^{\dagger} c_{k}$, is
\begin{equation}
P=\displaystyle\sum_{q>0}q(b_q^\dagger b_q-
\tilde{b}_q^\dagger \tilde{b}_q)+ \pi (\bar{d}_0+M/L)\,J.\\
\label{bosonP}
\end{equation}

In the coordinate-space formulation, the normalized
density field $\rho(x)$ is given
by $\rho(x)=\rho_R(x)+\rho_L(x)$:
\begin{equation}
\rho_R(x)=\frac{M}{2L}+
\displaystyle\sum_{q>0}\sqrt{q/2\pi L\lambda}
(e^{iqx}b_q+e^{-iqx}b_q^\dagger),
\label{rhofield}
\end{equation}
and $\rho_L(x)$ is similarly constructed from
$\tilde{b}_q$ and $\tilde{b}_q^\dagger$. The boson
field $\phi(x)$, which is conjugated to $\rho(x)$
and satisfies $[\phi(x),\rho(x')]=i\delta(x-x')$,
is $\phi(x)=\phi_R(x)+\phi_L(x)$ with
\begin{displaymath}
\phi_R(x)= \frac{\phi_0}{2}+\frac{\pi Jx}{2L}+i
\displaystyle \sum_{q>0}\sqrt{\pi\lambda/
2qL}(e^{iqx}b_q-e^{-iqx}b_q^\dagger),
\end{displaymath}
and a similar $\phi_{L}(x)$. Here $M$ and $J$ are operators
with integer eigenvalues, and $\phi_0$ is an angular variable
conjugated to M: $[\phi_0,M]=i$. The Hamiltonian (\ref{bosonH})
becomes
\begin{equation}
H_B =\frac{v_s}{2\pi}
\int_0^Ldx\; [\Pi(x)^2+(\partial_xX(x))^2],
\label{fieldH}
\end{equation}
where $\Pi(x)=\pi\lambda^{1/2}\rho(x)$ and
$X(x)=\lambda^{-1/2}\phi(x)$. With
$X(x,t)=e^{iHt}X(x)e^{-iHt}$, the
Lagrangian density reads
\begin{equation}
{\cal L}=\frac{v_s}{2\pi}\,\partial_\alpha
X(x,t)\,\partial^\alpha X(x,t).
\end{equation}

We recognize that $\cal{L}$ is the Lagrangian
of a $c=1$ CFT\cite{CFT}. Since $\phi_{0}$
is an angular variable, there is a hidden
invariance in the theory under $\phi\to\phi+2\pi$.
The field $X$ is thus said to be ``compactified''
on a circle, with a radius that is determined
by the exclusion statistics:
\begin{equation}
X\sim X+2\pi R,\;\;\; R^2=1/\lambda.
\end{equation}
States $V[X]|0\rangle$ or operators
$V[X]$ are {\it allowed} only if they respect
this invariance, so quantum numbers of
quasiparticles are strongly constrained. The
particle-hole duality\cite{BerWu,NaWil}, i.e.\
$\lambda\leftrightarrow 1/\lambda$ and
$M\leftrightarrow J$, in eq. (\ref{bosonH})
is just a duality $R\leftrightarrow 1/R$.
Moreover, the partition function of IEG
(in the low-$T$ limit) can be rewritten as
$Z=Tr_{\cal H}[q^{L^R_0} \bar{q}^{L^L_0}]$,
where $q=e^{iv_s/T}$. The zeroth generators
of the Virasoro algebra are $L^{R,L}_0
=v_s^{-1}H_{R,L}^{(b)}+\frac{\pi}{4L} [J R\mp M /R]^2$.
The constraint $M=J\,({\rm mod}\; 2)$ makes our
variant of $c=1$ CFT have an unusual spectrum and
duality relation\cite{CFT}.

We also see that the Hamiltonian (\ref{fieldH})
agrees with Haldane's harmonic fluid description
of Luttinger liquids\cite{Hald3}. Thus the critical
properties of IEG reproduce those of Luttinger
liquids. Or one may say that IEG can be used to
characterize the fixed points of Luttinger liquids.

Using bosonization techniques, the computation
of correlation functions and critical exponents
in IEG is similar to that for Luttinger
liquids\cite{Hald2,Hald3}. A careful construction
for the allowed boson field with charge-1
leads to
\begin{eqnarray}
\Psi^\dagger_B(x,t)=\rho(x)^{1/2}
&&\sum_{m=-\infty}^{\infty} e^{iO_m}
:e^{i(\lambda^{1/2} + 2m/\lambda^{1/2})X_{R}(x_-)}:
\nonumber \\
&& :e^{i(\lambda^{1/2}-2m/\lambda^{1/2})X_{L}(x_+)}:\; ,
\label{bosonPsi}
\end{eqnarray}
where the hermitian, constant-valued operators $O_m$
satisfy $[O_m, O_{m'}]=i\pi(m-m')$\cite{comm2}.
The excluson operator reads
$\Psi^\dagger_{\lambda}(x)=
:\Psi^\dagger_{B}(x)e^{i\lambda^{1/2}(X_R(x)-X_L(x))}:$,
obeying
$\Psi^\dagger_{\lambda}(x)\Psi^\dagger_{\lambda}(x')-
e^{i\pi\lambda sgn(x-x')}\Psi^\dagger_{\lambda}(x')
\Psi^\dagger_{\lambda}(x)=0$ for $x\not= x'$.
The multi-sector density operator for exclusons is
\begin{eqnarray}
&&\hat{\rho}(x)=\Psi^\dagger_{\lambda}(x)
\Psi_{\lambda}(x)=\Psi_B^\dagger(x)\Psi_B(x)\nonumber \\
&=& \rho(x)\sum{}_m :\exp\{i2m[X_R(x)-X_L(x)]/\lambda^{1/2}\}:
\end{eqnarray}
Dynamical correlation functions can be easily calculated:
\begin{eqnarray}
{\begin{array}{rcl}
\langle \hat{\rho}(x,t)\hat{\rho}(0,0)\rangle
&\approx&\bar{d}_0^2 \Biggl[1+\displaystyle
\frac{1}{(2\pi\bar{d}_0)^2\lambda}\Biggl(
\frac{1}{x_+^2}+\frac{1}{x_-^2}\Biggr)\nonumber \\
&+&{\displaystyle\sum_{m=1}^{\infty}} A_m
\frac{1}{[x_+x_-]^{m^2/\lambda}}
\cos(2\pi\bar{d}_0mx)\Biggr],\nonumber \\[2mm]
G(x,t;\lambda)&\equiv& \langle \Psi^\dagger_{\lambda}(x,t)
\Psi_{\lambda}(0,0)\rangle \\
\approx \bar{d}_0\displaystyle\sum_{m=-\infty}^{\infty}
&B_m&\frac{1}{x_-^{(m+\lambda)^2/\lambda}}
\frac{1}{x_+^{m^2/\lambda}}e^{i(2\pi(m+\lambda/2)x+\mu t)},
\end{array}}
\label{green}
\end{eqnarray}
with $A_m$ and $B_m$ regularization-dependent
constants. These correlation functions coincide with
the asymptotic ones\cite{Ha} in the
Calogero-Sutherland model.

The single-hole state, i.e.\  $\Psi^\dagger_{1/\lambda}
|0\rangle \equiv \Psi_\lambda (\lambda\to 1/\lambda)
|0\rangle $, with charge $-1/\lambda$ alone
is not allowed. The minimum allowed multi-hole state
is given by $\Psi^\dagger_{1/\lambda}(x_1)...
\Psi^\dagger_{1/\lambda}(x_p)|0\rangle$
if $\lambda=p/q$ is rational.
One may obtain, e.g.\ ,
\begin{equation}
\langle [\Psi^\dagger_{1/\lambda}(x,t)]^p
[\Psi_{1/\lambda}(0,0)]^p\rangle
\sim [G(x,t;1/\lambda)]^p.
\end{equation}
A more interesting allowed operator is what creates
$q$ particle excitations accompanied by $p$ hole
excitations: $\hat{n}(x,t)=[\Psi^\dagger_{\lambda}(x,t)]^q
[\Psi^\dagger_{1/\lambda}(x,t)]^p$.
We note the similarity of this operator
to Read's order parameter\cite{Read} for fractional
quantum Hall fluids (in bulk). Its correlation
function can be calculated by using Wick's theorem:
\begin{equation}
\langle \hat{n}(x,t)\hat{n}(0,0) \rangle\sim
[G(x,t;\lambda)]^q[G(x,t;1/\lambda)]^p.
\end{equation}
If the contribution from the $m=0$ sector dominates,
then one gets $\langle \hat{n}(x,t)\hat{n}(0,0)
\rangle \sim (x-v_st)^{-(p+q)}$.

Now we turn to discussing the effects of mutual
statistics. Consider a GIG with the
statistical matrix (\ref{stat}) given by
$g(k-k')=\delta(k-k')+\Phi(k-k')$. Here
$\Phi(k)=\Phi(-k)$ is a smooth function
and stands for mutual statistics between particles
with different momenta, in contrast to IEG,
for which $\Phi(k)$ is a $\delta$-function.
The thermodynamic properties of GIG is also given
by eq. (\ref{Omega}), but now $w(k,T)$ satisfies
instead an integral equation\cite{Wu,BerWu} which, in
terms of the dressed energy (\ref{dressE}), is
of the form
\begin{displaymath}
\epsilon(k,T)=\epsilon_0(k)+T
\int \frac{dk'}{2\pi}\;\Phi (k-k')\,
\ln (1+ e^{-\epsilon(k',T)/T})\; ,
\end{displaymath}
where $\epsilon_0(k)\equiv k^{2}$,
and we have shifted the dressed energy by
chemical potential $\mu$. At $T=0$, the Fermi momentum
$k_F$ is determined by $\epsilon(\pm k_F)=0$.
Introduce
\begin{eqnarray}
(\alpha\cdot\beta) [-k_{F},k_F] &\equiv& \int^{k_F}_{-k_F}
\frac{dk}{2\pi}\; \alpha (k)\, \beta (k)\, ,\\
(\Phi \cdot\alpha) (k;-k_{F},k_F]
&\equiv& \int_{-k_F}^{k_F} \frac{dk'}{2\pi}\;\Phi (k-k')
\,\alpha (k')\, .
\end{eqnarray}
Then both $\rho (k)$ and $\epsilon(k)$ in the ground
state satisfy an integral equation like
\begin{equation}
\alpha (k) = \alpha_0(k)-(\Phi\cdot \alpha )\, (k;-k_F,k_F]\;.
\label{inteq}
\end{equation}
The dressed momentum $p(k)$ is related to $\rho (k)$
by $p'(k)=2\pi\rho (k)$ and $p(k)=-p(-k)$. The ground state
energy is given by $E_0/L=(\epsilon_0\cdot \rho)[-k_F,k_{F}]
=(\epsilon\cdot\rho_0)[-k_F,k_{F}]$.
These equations are of the same form as those in
the thermodynamic Bethe ansatz \cite{YangYang},
hence the Luttinger-liquid relation $v_{s} =
\sqrt{v_{N}v_{J}}$ remains true\cite{Hald4}. A simple
proof can be sketched as follows. The sound velocity
is well-known: $v_s= \partial \epsilon(p_{F})/\partial p_{F}$.
The charge velocity is given by $v_N=v_s\, z(k_F)^{-2}$, where
the dressed charge $z(k)$ is given \cite{Hald4}
by the solution to the integral equation
$z(k)=1- (\Phi \cdot z) (k;-k_F,k_{F}]$.
This relation can be easily derived from the definitions
$v_N=L \partial \mu/\partial N_{0}$ and
$z(k)=-\delta \epsilon(k)/\delta \mu$.  To
create a persistent current, let us boost the Fermi
sea by $\pm k_{F} \to \pm k_F+\Delta$, where
$\Delta=z(k_F)/L\rho(k_F)$. Then the total energy
of the state with the persistent current is
\begin{eqnarray}
E_\Delta/L&=&(\epsilon_0\cdot\rho_\Delta)[-k_F
+\Delta,k_{F}+\Delta] \nonumber\\
&=& (\epsilon_\Delta\cdot\rho_0)[-k_F+\Delta,k_{F}+\Delta],
\end{eqnarray}
where $\rho_{\Delta} (k)=\rho_0(k)-
(\Phi\cdot \rho_\Delta)(k;-k_F+\Delta,k_{F}+\Delta]$ and
$\epsilon_\Delta(k)=\epsilon_0(k)-
(\Phi \cdot\epsilon_\Delta)(k;-k_F+\Delta,k_{F}+\Delta]$.
Now, using the last expression of $E_\Delta$,
it is easy to show $E_\Delta-E_0=L\Delta^2
\epsilon'(k_F)\rho (k_F)=(2\pi/L)v_s z(k_F)^{-2}$.
This verifies $v_J=v_s z(k_F)^2$. In view of
eq. (\ref{Velo}), at low energies
GIG looks like IEG with
\begin{equation}
\lambda_{eff}=z(k_{F})^{-2}.
\label{effstat}
\end{equation}

It can be shown that it is the effective statistics
(\ref{effstat}) that controls the low-$T$ critical
properties of GIG, as $\lambda$ does for IEG\@.
Linearization near the Fermi points and bosonization of
the low-energy effective Hamiltonian go pretty the same
way as before for IEG\@. The only difference now
is that the linearized dispersion for dressed
energy $\epsilon_\pm(k)=\pm \epsilon'(k_F)(k\mp k_F)+\mu
= \pm v_s(p(k)\mp p_F)+\mu$, is smooth at $k\sim \pm k_{F}$.
So bosonization is standard and the bosonized Hamiltonian
is the same as eq. (\ref{bosonH}) for IEG, only
with $\lambda$ replaced by $\lambda_{eff}$.
An (allowed) $\Psi_{\lambda_{eff}}^\dagger$ describes the
particle excitations near the Fermi surface with both
anyon and exlcusion statistics being $\lambda_{eff}$.
In this sense, one may say that the effect of
mutual statistics is to renormalize the statistics.

Here we remark that in IEG,
$\Phi(k,k')=(\lambda-1)\,\delta(k-k')$ is not smooth,
so the dressed charge has a jump at $k_F$:
$z(k_F^+)=1$ and $z(k_F^-)=\lambda^{-1}$ for
$k_F^\pm=k_F\pm 0^+$. The general Luttinger-liquid
relation is of the form
\begin{equation}
v_N=v_s[z(k_F^+)z(k_F^-)]^{-1},~~~v_J=v_sz(k_F^+)z(k_F^-).
\end{equation}

Finally we examine non-ideal gases, e.g., with general
Luttinger-type density-density interactions
\begin{equation}\begin{array}{rcl}
H_I=\displaystyle\frac{\pi}{L}\sum_{q\geq 0}
[U_{q}(\rho_q\rho_q^\dagger+\tilde{\rho}_q\tilde{\rho}_q^\dagger)
+ V_{q}(\rho_q\tilde{\rho}_q^\dagger+\tilde{\rho}_q\rho_q^\dagger)],
\end{array}\end{equation}
which is added to the Hamiltonian describing a GIG.
After bosonization, the total Hamiltonian remains
bilinear in densities, so it is trivial to diagonalize
it using the Bogoliubov transformation. The diagonalized
Hamiltonian is again of the harmonic-fluid form
(\ref{bosonH})
with $b_q$ and $\tilde{b}_q$ replaced by corresponding
operators for Bogoliubov quasiparticles, and with
the velocities renormalized:
$v_{s}\to\tilde{v}_s=|(v_s+U_0)^2-V_0^2|^{1/2}$,
$v_{N}\to \tilde{v}_N=\tilde{v}_s e^{-2\tilde{\varphi}_0}$,
and $v_{J}\to\tilde{v}_J=\tilde{v}_se^{2\tilde{\varphi}_0}$.
Thus, the Luttinger-liquid relation (\ref{Velo}) survives,
with $\lambda_{eff}$ of GIG renormalized to
\begin{equation}
\tilde{\lambda}_{eff} = e^{-2\tilde{\varphi_0}},\,
{\rm with}\; \tanh(2\tilde{\varphi}_0)
=\frac{v_J-v_N-2V_0}{v_J+v_N+2U_0}.
\end{equation}
Note that the new fixed point depends both on the
position of the Fermi points and on the interaction
parameters $U_{0}$ and $V_{0}$, leading to
``non-universal''exponents.

In passing we observe several additional implications
of this work: 1) Our bosonization and operator derivation
of CFT at low energies or in low-$T$ limit can be
applied to Bethe ansatz solvable models, including the
long-range (e.g. Calogero-Sutherland) one. 2) Here we
have only consider one-species cases, i.e. with excitations
having no internal quantum numbers such as spin. Our
bosonization and characterization of Luttinger liquids
are generalizable to GIG with multi-species, presumably
with the effective statistics matrix related to the dressed
charge matrix. 3) The chiral current algebra in eq. (\ref{density})
with $\lambda=1/m$ coincides with that derived by Wen\cite{Wen}
for edge states in $\nu=1/m$ fractional quantum Hall fluids.
So these edge states and their chiral Luttinger-liquid
fixed points can be described in terms of chiral IEG\@.

In conclusion, we have shown that 1D IEG without mutual
statistics can exactly reproduce the low-energy and
low-$T$ properties of (one-component) Luttinger liquids.
Moreover, mutual statistics and Luttinger-type
interactions in a GIG only shift the value of
$\lambda_{eff}$. Thus the essence of Luttinger liquids
is to have an IEG obeying FES as their fixed point.
It is conceivable that some strongly correlated
systems, exhibiting non-Fermi liquid behavior, in two or
higher dimensions can also be characterized as having
a GIG with appropriate statistics matrix as their
low-energy or low-temperature fixed point.

This work was supported in part by NSF grand PHY-93094598.

\end{document}